\documentclass{article}
\usepackage{algorithm} 
\usepackage{algpseudocode} 
\usepackage{spconf,amsmath,graphicx}
\usepackage[table,xcdraw]{xcolor}
\usepackage{booktabs} 
\usepackage{graphicx}
\usepackage{hyperref}
\usepackage[font=footnotesize]{subfig}

\graphicspath{ {./pictures/} }
\newcommand{\comment}[1]{}
\usepackage{amssymb}
\usepackage{xcolor}
\usepackage{hyperref}
\usepackage{booktabs}

\usepackage{xcolor}
\pagecolor{white}

\title{Small-Footprint Slimmable Networks for Keyword Spotting}

\author{Share\LaTeX}

\name{Zuhaib Akhtar$^{1,2}$, Mohammad Omar Khursheed$^{1}$, Dongsu Du$^{1}$, Yuzong Liu$^{1}$  \thanks{Zuhaib Akhtar performed this work during an internship at Amazon}}
\address{Alexa Perceptual Technologies, Amazon$^{1}$ New York University, NY$^{2}$ }
\begin{document}
%
\maketitle
\begin{abstract}
In this work, we present Slimmable Neural Networks applied to the problem of small-footprint keyword spotting. We show that slimmable neural networks allow us to create super-nets from Convolutional Neural Networks and Transformers, from which sub-networks of different sizes can be extracted. We demonstrate the usefulness of these models on in-house voice assistant data and Google Speech Commands, and focus our efforts on models for the on-device use case, limiting ourselves to less than 250k parameters. We show that slimmable models can match (and in some cases, outperform) models trained from scratch. Slimmable neural networks are therefore a class of models particularly useful when the same functionality is to be replicated at different memory and compute budgets, with different accuracy requirements.
%
\end{abstract}
\begin{keywords}
\emph{Slimmable Neural Networks, Transformers, Keyword Spotting, Convolutional Neural Networks}
\end{keywords}
\section{Introduction}
\label{sec:intro}
Keyword spotting, or wakeword detection \cite{szoke2005comparison}, is the act of identifying the presence of a keyword within a stream of audio. This can be thought of as a binary classification problem in the simplest case, detecting if there's a keyword or not. Wakeword detection has become the first point of interaction that a user has with voice assistant systems that are now ubiquitous in everyday life, such as Amazon's Alexa and Apples' Siri. With the prevalence of voice assistants in different types of devices such as earbuds, mobile devices, and smart speakers, there is a constant need to develop on-device keyword spotting models with tradeoffs between model accuracy and on-device resource constraints such as model size and CPU computation. As a result, researchers need to develop deep learning models with different resource constraints, which is expensive in compute and research time.

In recent work, there have been promising approaches towards developing deep learning models with different accuracy-resource tradeoffs for edge devices. The key idea is to train a single network that allows one to derive sub-networks with different tradeoffs between accuracy and resource constraint. One example is the class of networks called slimmable neural networks \cite{yu2018slimmable}, which train a neural network executable at different widths via weight-sharing and switchable Batch Normalization. Dynamic neural networks are another paradigm in which the network dynamically adapts its computation graph and parameters to different inputs and permits tradeoff between accuracy and inference efficiency \cite{han2021dynamic}. Another notable work Once-for-All (OFA) network was proposed in \cite{cai2020once}, which allows one to train one super-network once and derive multiple sub-networks with different resource contraint requirements. OFA also mitigates the large computational cost in conventional neural architecture search (NAS) by decoupling the network training and search.

In this paper, we propose the use of slimmable neural networks for small-footprint on-device keyword spotting. Unlike the original slimmable neural networks which were applied to moderate-size (3-25M parameters) networks, our focus is to explore the feasibility of training one small-footprint ($<$250k parameters) neural network to derive sub-networks that achieve tradeoffs between accuracy and device resource constraints. Our contributions are summarized as follows: 1) We propose lightweight CNN-based keyword spotting models using slimmable architectures; 2) We extend the original slimmable networks for keyword spotting with detailed design on slimmable self-attention module; 3) We demonstrate the effectiveness of the proposed method on both a large-scale in-house voice assistant dataset and the public Google Speech Commands dataset \cite{https://doi.org/10.48550/arxiv.1804.03209}. To our knowledge, this is the first work to apply the slimmable networks to multiple architectures including CNN and Transformers on a large-scale keyword spotting dataset under tight on-device resource constraints (10k-250k parameters).

\section{Related work}

Keyword spotting models \cite{chen2014small,sainath15b_interspeech,panchapagesan2016multi,sun2017compressed,khursheed2021tiny} are deployed on-device and typically use a small-footprint deep learning model with minimal latency. With the growing need to deploy deep learning applications to edge devices, it takes significant manual efforts from researchers to develop models to accommodate different on-device budget constraints. There is a plethora of work on developing one single super-network with multiple sub-networks that can be executed on their own and meet the on-device computation budget, including slimmable networks \cite{yu2018slimmable,yu2019universally,yu2019autoslim}, dynamic neural network \cite{liu2018dynamic,huang2017multi,han2021dynamic}, once-for-all network training \cite{cai2020once}. More recently in speech processing applications, there has been further work exploring the feasibility of this concept. In  \cite{wu2021dynamic}, a dynamic sparsity neural network (DSNN) was proposed for speech recognition. The idea of DSNN is to train one single over-parameterized super-network and apply pruning masks defined from a fixed set of sparsity ratio values. In \cite{yang2022omni}, an Omni-sparsity DNN was proposed for on-device speech recognition. Similar to \cite{wu2021dynamic}, layer-wise pruning masks were applied to sample sub-networks with different sparsity ratios during super-network training. After the super-network has been trained, an evolutionary searching algorithm was adopted to find the optimal sub-network for any sparsity requirement. In speech representation pretraining \cite{wang2022lighthubert}, an OFA \cite{cai2020once} was applied to find the optimal sub-network in a computationally-intensive HuBERT model. A closely related idea is that of Neural Architecture Search \cite{elsken2019neural} \cite{mehrotra2020bench} which has been used for the problem of speech recognition.

\section{Proposed Approach}
 We show that Convolutional Neural Networks and Transformer models are slimmable via the general procedure described in \textbf{Algorithm 1}. Slimmable training works by defining a width list over which the network will run. During training, the larger network is set to a particular width which is used as a normal neural network to get predictions over a batch, following which losses and gradients are computed. The gradients are saved and weights of the network are not changed at this stage. Following this, the same batch of data is passed again to a model with a smaller width, with the parameters shared from the first larger model. This is done till all the gradients are calculated for every sub-network. Finally, all the gradients are summed up and the weights are updated for the full network.
\begin{algorithm}
	\begin{algorithmic}[1]
	    \State ${width\_list} = [1.0, 0.75, 0.50, 0.25]$
	    \State ${model} = get\_model(width\_list)$
	    \State ${grad\_list} = []$
		\For {$batch \in \{dataset\}$}
			\For {$width \in \{width\_list\}$}
			    \State $slim\_network(width)$
                \State $predictions = model(batch)$
                \State $loss = loss\_function(predictions, labels)$
                \State $gradient = get\_gradients(loss, model.weigths)$
                \State $grad\_list.append(gradient)$
            \EndFor
			\State $overall\_gradients = sum\_gradient(grad\_list)$
			\State $apply\_gradients(model, overall\_gradients)$
		\EndFor
	\end{algorithmic}
	\caption{Slimmable Network Training} 
	\label{algo}
\end{algorithm}

\subsection{Slimming CNNs} 
Slimming in a convolution neural network works by sharing weights between different widths. As the weights are kernels in a convolution layer, when the network switches to a different width, a corresponding number of kernels are dropped to reduce the weight of the network according to the width the network has been set. As the intermediate tensor output is changing between different layers during slimming, the size of individual kernels are also reduced for the convolutional layer. Both of these techniques to reduce kernels makes slimming of the convolution possible. We also used switchable BatchNorm which assigns a separate BatchNorm layer \cite{yu2018slimmable} for each width. \textbf{Fig. 1} shows slimming of convolution neural network by our approach.

\begin{figure}[h]
    \centering
    \includegraphics[width=8.5cm]{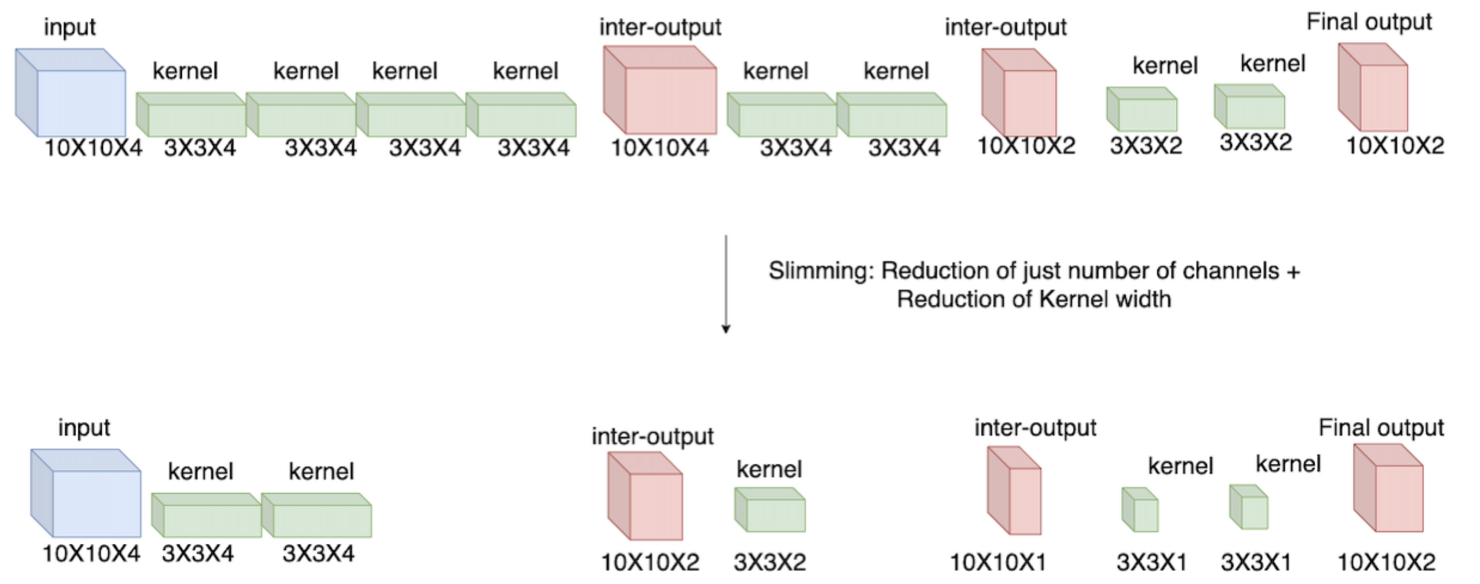}
    \caption{\emph{The figure shows of an example how end-to-end slimming is done for a convolutional neural network. For the first layer, kernels are dropped. For intermediate layers, kernels are dropped as well as reduced in width. For the last layer, kernels are reduced in width.}}
    \label{fig:cn}
\end{figure}

\subsection{Slimming Transformers}  For slimming of transformers \cite{https://doi.org/10.48550/arxiv.1706.03762}, all the dense layers are slimmed within every transformer block.  Each intermediate input gets reduced to a dimension determined by the width of the sub-network under consideration via a slimmable dense layer. During the slimming of dense layers in intermediate sub-blocks, a fraction of weights are switched off according to the width, and this is followed by the usual operations while taking the reduced dimension into account. A dense layer at the beginning of each block projects the input to a lower dimension which makes the output compatible with the residual connection. The final dense layer of the transformer ensures that it always produces output of same dimension (number of classes) irrespective of input dimension (which are reduced during slimming). The slimmable transformer is designed in such a way that LayerNorm  acts as a switch. For each width, a separate LayerNorm layer is used such that a particular layer that is assigned to that width is selected at run-time. This is implemented in a fashion similar to the switchable BatchNorm layer in CNNs.\\
\textbf{Equation 1, 2 and 3} shows the working of slimmable attention. Query (Q), Key (K) and Value (V) tensors are of different sizes according to the width of the transformer sub-network. Attention operation such as dot-product takes place on a lower dimension for the sub-networks according to the width. This makes lower-width models compute and memory efficient. These lower-dimension tensors are generated by slimming down Q, K and V weight matrices. \textbf{Fig. 2} shows the slimming of a transformer by our approach.
\begin{equation}
Slim\_Attn(Q, K, V) = softmax(\frac{Q[:i]K[:i]^T}{\sqrt{d_k[:i]}})V[:i]\\
\end{equation}
\begin{equation}
i\in [d_k*width[0], d_k*width[1],\ldots,d_k*width[n]]\\
\end{equation}
\begin{equation}
width\in [1, (n-1)/n, (n-2)/n,\ldots, (1/n)], n>1 (integer)
\end{equation}

\begin{figure}[h]
    \centering
    \includegraphics[width=5cm]{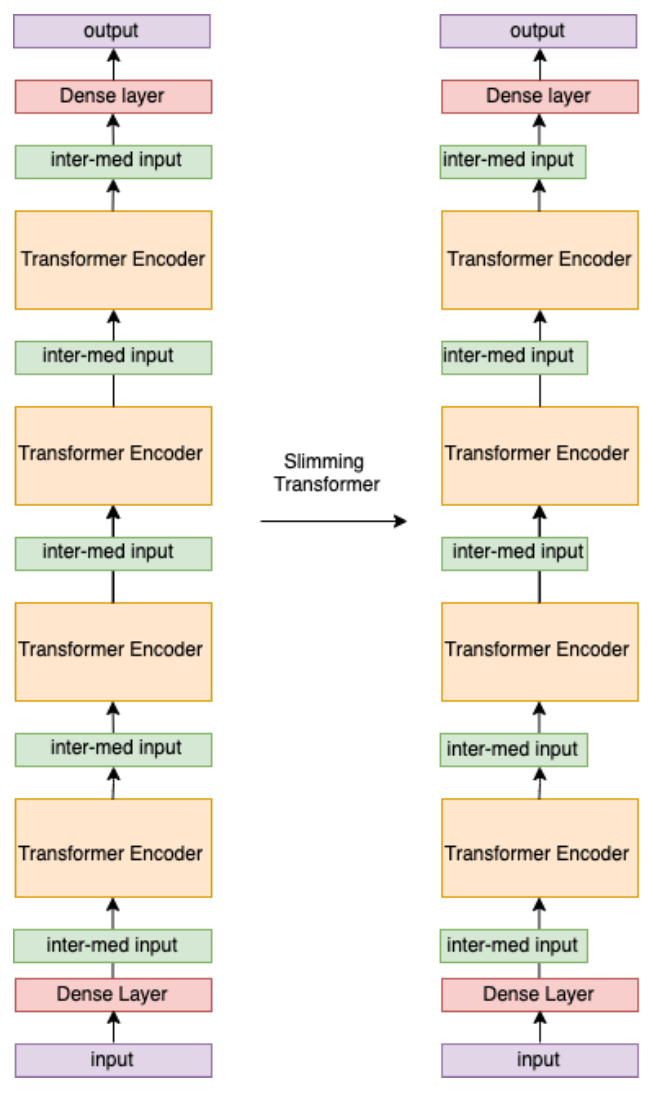}
    \caption{\emph{The figure shows of an example how end-to-end slimming is done for transformers.}}
    \label{fig:galaxy}
\end{figure}
\begin{table}[!htpb]
\centering
\scalebox{0.8}{\begin{tabular}{
>{\columncolor[HTML]{FFFFFF}}c 
>{\columncolor[HTML]{FFFFFF}}c 
>{\columncolor[HTML]{FFFFFF}}c 
>{\columncolor[HTML]{FFFFFF}}c 
>{\columncolor[HTML]{FFFFFF}}c }
\toprule
{\color[HTML]{000000} \textbf{CNN Layers}} & {\color[HTML]{000000} \textbf{Kernel size}} & {\color[HTML]{000000} \textbf{Channel}} & {\color[HTML]{000000} \textbf{Striding}} & {\color[HTML]{000000} \textbf{Pooling}} \\
\toprule
{\color[HTML]{000000} 1}                                              & {\color[HTML]{000000} (5, 4)}       & {\color[HTML]{000000} 32}   & {\color[HTML]{000000} (1, 2)}           & {\color[HTML]{000000} (2, 1)}               \\
{\color[HTML]{000000} 2}                                              & {\color[HTML]{000000} (3, 4)}       & {\color[HTML]{000000} 32}   & {\color[HTML]{000000} (1, 3)}           & {\color[HTML]{000000} (2, 1)}                \\
{\color[HTML]{000000} 3}                                              & {\color[HTML]{000000} (4, 4)}       & {\color[HTML]{000000} 40}   & {\color[HTML]{000000} (1, 2)}           & {\color[HTML]{000000} (2, 1)}              \\
{\color[HTML]{000000} 4}                                              & {\color[HTML]{000000} (7, 4)}       & {\color[HTML]{000000} 128}   & {\color[HTML]{000000} (1, 1)}           & {\color[HTML]{000000} (1, 1)}              \\
{\color[HTML]{000000} 5}                                              & {\color[HTML]{000000} (1, 1)}       & {\color[HTML]{000000} 160}   & {\color[HTML]{000000} (1, 1)}           & {\color[HTML]{000000} (1, 1)}               \\
\toprule
{\color[HTML]{000000} \textbf{Transformer model}} & {\color[HTML]{000000} \textbf{dim}} & {\color[HTML]{000000} \textbf{mlp-dim}} & {\color[HTML]{000000} \textbf{heads}} & {\color[HTML]{000000} \textbf{layer}} \\
\toprule
{\color[HTML]{000000} Transformer (in-house data)}                                              & {\color[HTML]{000000} 64}       & {\color[HTML]{000000} 128}  & {\color[HTML]{000000} 1}           & {\color[HTML]{000000} 3}               \\
{\color[HTML]{000000} Transformer-GoogleSpeech}                                              & {\color[HTML]{000000} 64}       & {\color[HTML]{000000} 64}   & {\color[HTML]{000000} 1}           & {\color[HTML]{000000} 2}                 \\

\end{tabular}}
\caption{\emph{CNN and Transformer Architectures}}
\end{table}





\section{Experiments and Results}


\begin{table}[!htpb]
\scalebox{0.75}{\begin{tabular}{
>{\columncolor[HTML]{FFFFFF}}c 
>{\columncolor[HTML]{FFFFFF}}c 
>{\columncolor[HTML]{FFFFFF}}c 
>{\columncolor[HTML]{FFFFFF}}c 
>{\columncolor[HTML]{FFFFFF}}c }
\toprule
{\color[HTML]{000000} \textbf{Model Name}} & {\color[HTML]{000000} \textbf{Relative False Accepts}} & {\color[HTML]{000000} \textbf{Parameters}} & {\color[HTML]{000000} \textbf{Multiplies}} \\
\toprule
{\color[HTML]{000000} CNN-Scratch-Width-1}                                              & {\color[HTML]{000000} Baseline (1)}       & {\color[HTML]{000000} 199k}           & {\color[HTML]{000000} 3.5M}               \\
{\color[HTML]{000000} CNN-Scratch-Width-0.75}                                           & {\color[HTML]{000000} 1.11}               & {\color[HTML]{000000} 122k}            & {\color[HTML]{000000} 2.1M}                \\
{\color[HTML]{000000} CNN-Scratch-Width-0.5}                                            & {\color[HTML]{000000} 1.28}               & {\color[HTML]{000000} 50k}            & {\color[HTML]{000000} 1.1M}              \\
{\color[HTML]{000000} CNN-Scratch-Width-0.25}                                           & {\color[HTML]{000000} 1.78}               & {\color[HTML]{000000} 13k}            & {\color[HTML]{000000} 0.35M}              \\
\midrule
{\color[HTML]{000000} CNN-Slim-Width-1}                                              & {\color[HTML]{000000} 1.05}       & {\color[HTML]{000000} 199k}           & {\color[HTML]{000000} 3.5M}               \\
{\color[HTML]{000000} CNN-Slim-Width-0.75}                                           & {\color[HTML]{000000} 1.20}               & {\color[HTML]{000000} 122k}            & {\color[HTML]{000000} 2.1M}                \\
{\color[HTML]{000000} CNN-Slim-Width-0.5}                                            & {\color[HTML]{000000} 1.35}               & {\color[HTML]{000000} 50k}            & {\color[HTML]{000000} 1.1M}              \\
{\color[HTML]{000000} CNN-Slim-Width-0.25}                                 & {\color[HTML]{000000} 1.72}       & {\color[HTML]{000000} 13k}            & {\color[HTML]{000000} 0.35M}              \\
\end{tabular}}
\caption{\emph{CNN Results on in-house Dataset}}
\end{table}
\begin{table}[!htpb]
\scalebox{0.8}{\begin{tabular}{
>{\columncolor[HTML]{FFFFFF}}c 
>{\columncolor[HTML]{FFFFFF}}c 
>{\columncolor[HTML]{FFFFFF}}c 
>{\columncolor[HTML]{FFFFFF}}c 
>{\columncolor[HTML]{FFFFFF}}c }
\toprule
{\color[HTML]{000000} \textbf{Model Name}} & {\color[HTML]{000000} \textbf{Accuracy}} & {\color[HTML]{000000} \textbf{Parameters}} & {\color[HTML]{000000} \textbf{Multiplies}} \\
\toprule
{\color[HTML]{000000} CNN-Scratch-Width-1}                                              & {\color[HTML]{000000} 0.8915}       & {\color[HTML]{000000} 243k}           & {\color[HTML]{000000} 4.8M}               \\
{\color[HTML]{000000} CNN-Scratch-Width-0.75}                                           & {\color[HTML]{000000} 0.8878}               & {\color[HTML]{000000} 138k}            & {\color[HTML]{000000} 2.9M}                \\
{\color[HTML]{000000} CNN-Scratch-Width-0.5}                                            & {\color[HTML]{000000} 0.8636}               & {\color[HTML]{000000} 62k}            & {\color[HTML]{000000} 1.4M}              \\
{\color[HTML]{000000} CNN-Scratch-Width-0.25}                                           & {\color[HTML]{000000} 0.8455}               & {\color[HTML]{000000} 16k}            & {\color[HTML]{000000} 0.48M}              \\
\midrule
{\color[HTML]{000000} CNN-Slim-Width-1}                                              & {\color[HTML]{000000} 0.9047}       & {\color[HTML]{000000} 243k}           & {\color[HTML]{000000} 4.8M}               \\
{\color[HTML]{000000} CNN-Slim-Width-0.75}                                           & {\color[HTML]{000000} 0.8967}               & {\color[HTML]{000000} 138k}            & {\color[HTML]{000000} 2.9M}                \\
{\color[HTML]{000000} CNN-Slim-Width-0.5}                                            & {\color[HTML]{000000} 0.8726}               & {\color[HTML]{000000} 62k}            & {\color[HTML]{000000} 1.4M}              \\
{\color[HTML]{000000} CNN-Slim-Width-0.25}                                           & {\color[HTML]{000000} 0.8446}               & {\color[HTML]{000000} 16k}            & {\color[HTML]{000000} 0.48M}              \\
{\color[HTML]{000000} Res26}                                           & {\color[HTML]{000000} 0.95}               & {\color[HTML]{000000} 438k}            & {\color[HTML]{000000} N/A}              \\
\end{tabular}}
\caption{\emph{CNN Results on Google Speech Commands Dataset}}
\end{table}
\subsection{Data and Baseline models}
We trained all our models on 2 different tasks. The first task was to detect a unique keyword in our in-house voice assistant dataset, and the second is the Google Speech Commands 35-class classification task using TensorFlow \cite{https://doi.org/10.48550/arxiv.1603.04467}.
\\We use 76 input frames for our CNN models and 182 frames for transformer models, computing Log Mel Filter Bank Energies (LFBEs) every 10 ms over a window of 25 ms for our in-house dataset. For Google Speech Commands data, we use 98 frames of audio for both model types. We align the wakeword using a DNN-HMM based spotter model that places the keyword in the center of the context window being fed into the model for training. We train our models on over 20000 hours of de-identified in-house data and the full training set from Google Speech Commands, and evaluate on 100 hours for in-house data and the Google Speech Commands test set for the latter. Our baseline CNN architecture is a 5-layer CNN with 199k parameters (\textbf{Table 1}) for both in-house and Google Speech Commands datasets. Our baseline transformer architecture is a smaller version of the KWT \cite{berg2021keyword} architecture with 120k parameters for in-house data and 67k parameters for Google Speech Commands (\textbf{Table 1}).
For all experiments on the in-house dataset, we report relative improvement over a baseline model in terms of False Accepts at a fixed miss rate, and we provide absolute accuracy for google speech commands. We also report parameter counts and multiplication operations, since these provide a sense of the memory and compute requirements of our keyword spotting models. 

\subsection{Slimmable CNNs and Transformers}
When slimming a 199k parameter CNN to 4 different widths (1, 0.75, 0.5 and 0.25), we find an accuracy and compute trade-off. While the slimmed models are slightly worse compared to single models of the same size trained from scratch on our in-house  dataset as shown in \textbf{Table 2}, the training time is reduced by a factor of 4 per step, which is a significant savings in spite of the extra epochs (20 vs 30) it takes to improve slimmable models when compared to their counterparts trained from scratch. However, for the smallest width (0.25), the slimmable model outperforms the equivalent model trained from scratch. \\We also see that training time for multiple widths does not scale linearly. Instead, as \textbf{Table 4} shows, there is only a 3.72x increase in training time when the number of widths goes from 1 to 40. This points to the possibility of training a single super-network, from which sub-networks of a variety of sizes can be extracted, in over 10 times less time than individually training each width from scratch.

\begin{table}[!htpb]
\centering
\scalebox{0.8}{\begin{tabular}{
>{\columncolor[HTML]{FFFFFF}}c 
>{\columncolor[HTML]{FFFFFF}}c  }
\toprule
{\color[HTML]{000000} \textbf{Total Widths}} & {\color[HTML]{000000} \textbf{Training Time Per Step (sec)}}\\
\toprule
{\color[HTML]{000000} 1}                                              & {\color[HTML]{000000} 1.01}       \\
{\color[HTML]{000000} 2}                                           & {\color[HTML]{000000} 1.53}        \\
{\color[HTML]{000000} 3}                                            & {\color[HTML]{000000} 1.60}               \\
{\color[HTML]{000000} 4}                                           & {\color[HTML]{000000} 1.63}               \\
{\color[HTML]{000000} 5}                                              & {\color[HTML]{000000} 1.69}       \\
{\color[HTML]{000000} 10}                                           & {\color[HTML]{000000} 2.01}               \\
{\color[HTML]{000000} 20}                                            & {\color[HTML]{000000} 2.29}               \\
{\color[HTML]{000000} 40}                                           & {\color[HTML]{000000} 3.72}               \\
\end{tabular}}
\caption{\emph{Slimmable NNs Training Time}}
\end{table}
When slimming down a 120k parameter transformer model to the same 4 widths, we see a similar pattern emerge, with the slimmable models being slightly worse than those trained from scratch on the in-house dataset as shown in \textbf{Table 5}. 
\\ When we repeat this experiment on Google Speech Commands data (35 classes), we see that the slimmable models are actually slightly better in terms of accuracy than models trained from scratch, for all CNNs except the smallest, as shown in \textbf{Table 3}. This indicates that the slimmable NN has a gain of performance because of the inherent distillation present in the model due to the shared weights \cite{yu2018slimmable} across different widths. For the slimmable transformer, this pattern also holds true (\textbf{Table 6})

\begin{table}[!htpb]
\scalebox{0.7}{\begin{tabular}{
>{\columncolor[HTML]{FFFFFF}}c 
>{\columncolor[HTML]{FFFFFF}}c 
>{\columncolor[HTML]{FFFFFF}}c 
>{\columncolor[HTML]{FFFFFF}}c 
>{\columncolor[HTML]{FFFFFF}}c }
\toprule
{\color[HTML]{000000} \textbf{Model Name}} & {\color[HTML]{000000} \textbf{Relative False Accepts}} & {\color[HTML]{000000} \textbf{Parameters}} & {\color[HTML]{000000} \textbf{Multiplies}} \\
\toprule
{\color[HTML]{000000} Transformer-Scratch-Width-1}                                              & {\color[HTML]{000000} Baseline (1)}       & {\color[HTML]{000000} 120k}           & {\color[HTML]{000000} 18M}               \\
{\color[HTML]{000000} Transformer-Scratch-Width-0.75}                                           & {\color[HTML]{000000} 1.08}               & {\color[HTML]{000000} 76k}            & {\color[HTML]{000000} 10.1M}                \\
{\color[HTML]{000000} Transformer-Scratch-Width-0.5}                                            & {\color[HTML]{000000} 1.15}               & {\color[HTML]{000000} 43k}            & {\color[HTML]{000000} 4.5M}              \\
{\color[HTML]{000000} Transformer-Scratch-Width-0.25}                                           & {\color[HTML]{000000} 1.36}               & {\color[HTML]{000000} 24k}            & {\color[HTML]{000000} 1.1M}              \\
\midrule
{\color[HTML]{000000} Transformer-Slim-Width-1}                                              & {\color[HTML]{000000} 1.19}       & {\color[HTML]{000000} 120k}           & {\color[HTML]{000000} 18M}               \\
{\color[HTML]{000000} Transformer-Slim-Width-0.75}                                           & {\color[HTML]{000000} 1.19}               & {\color[HTML]{000000} 76k}            & {\color[HTML]{000000} 10.1M}                \\
{\color[HTML]{000000} Transformer-Slim-Width-0.5}                                            & {\color[HTML]{000000} 1.27}               & {\color[HTML]{000000} 43k}            & {\color[HTML]{000000} 4.5M}              \\
{\color[HTML]{000000} Transformer-Slim-Width-0.25}                                            & {\color[HTML]{000000} 1.34}               & {\color[HTML]{000000} 24k}            & {\color[HTML]{000000} 1.1M}              \\
\end{tabular}}
\caption{\emph{Transformer Results on in-house dataset}}
\end{table}
\begin{table}[!htpb]
\scalebox{0.8}{\begin{tabular}{
>{\columncolor[HTML]{FFFFFF}}c 
>{\columncolor[HTML]{FFFFFF}}c 
>{\columncolor[HTML]{FFFFFF}}c 
>{\columncolor[HTML]{FFFFFF}}c 
>{\columncolor[HTML]{FFFFFF}}c }
\toprule
{\color[HTML]{000000} \textbf{Model Name}} & {\color[HTML]{000000} \textbf{\% Accuracy}} & {\color[HTML]{000000} \textbf{Parameters}} & {\color[HTML]{000000} \textbf{Multiplies}} \\
\toprule
{\color[HTML]{000000} Transformer-Scratch-Width-1}                                              & {\color[HTML]{000000} 0.8623}       & {\color[HTML]{000000} 67k}           & {\color[HTML]{000000} 7.2M}               \\
{\color[HTML]{000000} Transformer-Scratch-Width-0.75}                                           & {\color[HTML]{000000} 0.8463}               & {\color[HTML]{000000} 44k}            & {\color[HTML]{000000} 4.5M}                \\
{\color[HTML]{000000} Transformer-Scratch-Width-0.5}                                            & {\color[HTML]{000000} 0.8467}               & {\color[HTML]{000000} 26k}            & {\color[HTML]{000000} 2.4M}              \\
{\color[HTML]{000000} Transformer-Scratch-Width-0.25}                                           & {\color[HTML]{000000} 0.7295}               & {\color[HTML]{000000} 15k}            & {\color[HTML]{000000} 0.9M}              \\
\midrule
{\color[HTML]{000000} Transformer-Slim-Width-1}                                              & {\color[HTML]{000000} 0.8787}       & {\color[HTML]{000000} 67k}           & {\color[HTML]{000000} 7.2M}               \\
{\color[HTML]{000000} Transformer-Slim-Width-0.75}                                           & {\color[HTML]{000000} 0.8766}               & {\color[HTML]{000000} 44k}            & {\color[HTML]{000000} 4.5M}                \\
{\color[HTML]{000000} Transformer-Slim-Width-0.5}                                            & {\color[HTML]{000000} 0.8588}               & {\color[HTML]{000000} 26k}            & {\color[HTML]{000000} 2.4M}              \\
{\color[HTML]{000000} Transformer-Slim-Width-0.25}                                            & {\color[HTML]{000000} 0.7840}               & {\color[HTML]{000000} 15k}            & {\color[HTML]{000000} 0.9M}              \\
\end{tabular}}
\caption{\emph{Transformer Results on Google Speech Commands Dataset}}
\end{table}

\section{Conclusion and future work}
In this work, we show that slimmable networks are a viable option for small-footprint keyword spotting. We show that this method works across different types of architectures. Our work shows that slimmable NNs generalize to different datasets and situations. We explore the possibility of training multiple models at different budgets at once, and show, through analysis of training times, that slimmable NNs are several times more efficient when compared to training individual models from scratch. 
\\ The use of slimmable models can be used to test different sizes of a single architecture, and find the optimal network for a certain dataset. When one trains a supernetwork of a number of widths and evaluates on each of the sub-networks, it can be used to determine the minimum capacity of a model required to do well on that dataset. Considering that inference is usually a lot cheaper and faster to perform than training, this could enable models to be released on a variety of platforms with different memory, compute and performance requirements. Many companies that provide keyword spotting solutions as part of an AI assistant. Using slimmable models would allow for training and deploying models at scale. Since these AI assistants run on a variety of platforms, creating a single model that can run at different budgets (or have different budget sub-models extracted from itself) is a driver of efficiently deploying models across products, which vary from smartphones to earbuds to TVs. 
\\ We hope to extend this work by using more efficient automated slimming techniques such as AutoSlim \cite{yu2019autoslim}, and extending slimmable models to include novel architectures, such as RNNs. There also exists the possibility to slim the depth of a network, rather than just the width. For the purposes of edge computing, profiling slimmable models on various chipsets used for edge computing could give us an improved sense of the memory and compute of these models. 
\bibliographystyle{IEEEbib}
{\small{\bibliography{refs.bib}}
\end{document}